\documentstyle[12pt,epsfig,graphics,cite,amssymb,amsmath,bm]{article}
\begin{document}\bibliographystyle{plain}\begin{titlepage}
\renewcommand{\thefootnote}{\fnsymbol{footnote}}\hfill
\begin{tabular}{l}HEPHY-PUB 932/14\\UWThPh-2014-04\\March
2014\end{tabular}\\[2cm]\Large\begin{center}{\bf THE SPINLESS
RELATIVISTIC\\ WOODS--SAXON PROBLEM}\\[1cm]\large{\bf Wolfgang
LUCHA\footnote[1]{\normalsize\ \emph{E-mail address\/}:
wolfgang.lucha@oeaw.ac.at}}\\[.3cm]\normalsize Institute for High
Energy Physics,\\Austrian Academy of Sciences,\\Nikolsdorfergasse
18, A-1050 Vienna, Austria\\[1cm]\large{\bf Franz
F.~SCH\"OBERL\footnote[2]{\normalsize\ \emph{E-mail address\/}:
franz.schoeberl@univie.ac.at}}\\[.3cm]\normalsize Faculty of
Physics, University of Vienna,\\Boltzmanngasse 5, A-1090 Vienna,
Austria\\[2cm]{\normalsize\bf Abstract}\end{center}\normalsize

\noindent Motivated by the observation of a recent renewal of
rather strong interest in the description of bound states by
(semi-)relativistic equations of motion, we revisit, for the
example of the Woods--Saxon interactions, the eigenvalue problem
posed by the spinless Salpeter equation and recall various
elementary knowledge, considerations, and techniques that
practitioners seeking solutions to this specific reduction of the
Bethe--Salpeter equation~may find helpful.\vspace{3ex}

\noindent\emph{PACS numbers\/}: 03.65.Pm, 03.65.Ge, 12.39.Pn,
11.10.St\vspace{3ex}

\noindent\emph{Keywords\/}: relativistic bound states,
Bethe--Salpeter formalism, spinless Salpeter equation,
Rayleigh--Ritz variational technique, Woods--Saxon potential

\renewcommand{\thefootnote}{\arabic{footnote}}\end{titlepage}

\section{Introduction: Semirelativistic Equations of Motion}The
recent couple of years have witnessed a huge revival of interest
in investigating systems of (bound-state) constituents by means of
the semirelativistic equation of motion called the \emph{spinless
Salpeter equation\/},\footnote{Several aspects and facets of the
spinless Salpeter formalism are concisely reviewed in,
\emph{e.g.}, Refs.~\cite{Lucha92,Lucha94:Como,Lucha04:TWR}.} which
can be looked at from two rather opposite points~of~view:
\begin{itemize}\item On the one hand, it may be regarded as both
straightforward and very likely simplest generalization of the
nonrelativistic Schr\"odinger equation towards proper inclusion of
relativistic kinematics by the relativistically correct expression
for the kinetic~energy.\item On the other hand, it is encountered
along the course of \emph{three-dimensional reductions\/} of the
\emph{Bethe--Salpeter formalism\/} allowing for relativistically
covariant descriptions of bound states within quantum field theory
\cite{BSEa,BSEb,BSEc} if assuming instantaneous interaction
between and free propagation of the bound-state constituents
\cite{SE} and disregarding all negative-energy contributions
(which may be justified for semirelativistic and weakly bound
heavy constituents) as well as the spin degrees of freedom of
the~constituents.\footnote{The most essential steps of this
well-defined nonrelativistic reduction are sketched in,
\emph{e.g.}, Refs.~\cite{Lucha91,Lucha99:talksa,Lucha99:talksb}.}
\end{itemize}Irrespective of its conceptual roots, the spinless
Salpeter equation is an eigenvalue equation
$$H\,|\chi_k\rangle=E_k\,|\chi_k\rangle\ ,\qquad k=0,1,2,\dots\
,$$for the \emph{eigenstates\/} $|\chi_k\rangle$ and corresponding
\emph{eigenvalues\/} $E_k$ of a Hamiltonian $H$ composed of the
kinetic energies of the spin-zero particles forming the system
under consideration~and~a static potential subsuming all the
interactions of these constituents. The relativistic kinetic
energy of a spin-zero particle of effective mass $m$ and spatial
momentum $\bm{p}$ is represented~by the notorious
\emph{square-root operator\/}
$\sqrt{\bm{p}^2+m^2}.$\footnote{Throughout our analysis, we adopt,
of course, natural units appropriate for particle physics:
$\hbar=c=1.$} For systems consisting of just two particles, only
their relative momentum $\bm{p}$ and relative coordinate $\bm{x}$
are relevant for the Hamiltonian:\begin{equation}H\equiv
T(\bm{p})+V(\bm{x})\ .\label{Eq:H0}\end{equation}For simplicity,
let these two particles have equal masses $m$ and thus equal
kinetic energies:\footnote{The straightforward generalization to
the case of constituents with unequal masses is a trivial
exercise.}\begin{equation}H\equiv2\,\sqrt{\bm{p}^2+m^2}
+V(\bm{x})\ .\label{Eq:H}\end{equation}The, in general, nonlocal
nature of this operator renders difficult to arrive at exact
analytic statements about the solutions of its eigenvalue
equation. On the other hand, it is very easy to analyze the
reliability of any such solution, irrespective of how it has been
gained, by the relativistic generalization of the virial theorem
\cite{Lucha:RVTa,Lucha:RVTb} or to infer upper bounds~on~energies.

Here, we focus to spherically symmetric central potentials,
depending only on the radial coordinate $r\equiv|\bm{x}|,$ that
is, $V(\bm{x})=V(r).$ One particularly prominent interaction
potential is, because of its pivotal r\^ole for the shell model of
nuclear physics, the (real) Woods--Saxon potential, a short-ranged
potential, specified by three parameters: the coupling strength
$V_0$ determining the potential's depth, the potential's width
$R,$ and the surface~thickness~$a$~\cite{WSP}:\begin{equation}
V(r)=-\frac{V_0}{1+\exp\!\left(\frac{r-R}{a}\right)}\ ,\qquad
V_0>0\ ,\qquad R\ge0\ ,\qquad a>0\ .\label{Eq:WSP}\end{equation}
The Hamiltonian operator (\ref{Eq:H}) with the potential
(\ref{Eq:WSP}) poses what we call the semirelativistic spinless
Woods--Saxon problem. Very well aware of the difficulties
encountered when trying to derive \emph{exact\/} analytical
solutions to any such kind of problems, we formulate, by applying
standard tools, the trivial boundary conditions to all
corresponding \emph{approximate\/} solutions.

\section{Rigorous Upper Limit on Number of Bound States}
\label{Sec:NBS}One paramount specific of all bound-state problems
is the number of possible bound~states.

\subsection{Nonrelativistic Kinematics: the Schr\"odinger Equation}
In the case of the Schr\"odinger equation, results are abundant.
For the Schr\"odinger operator$$H=\frac{\bm{p}^2}{2\,\mu}+V(r)\
,\qquad\mu>0\ ,$$Bargmann \cite{Bargmann} found, for the number
$n_\ell$ of bound states with orbital angular momentum~$\ell,$
$$(2\,\ell+1)\,n_\ell\lneqq I\equiv2\,\mu\int_0^\infty{\rm
d}r\,r\,|V_-(r)|\ ,\qquad V_-(r)\equiv-\max[0,-V(r)]\ .$$Then,
$n_\ell\ge1$ implies $\ell\le\ell_{\rm max}=(I-1)/2,$ and the
total number $N$ of bound states~satisfies\begin{equation}
N=\sum_{\ell=0}^{\ell_{\rm max}}(2\,\ell+1)\,n_\ell\lneqq
I\,(\ell_{\rm max}+1)=\frac{I\,(I+1)}{2}\ .\label{Eq:BB}
\end{equation}

\subsection{Relativistic Kinematics: the Spinless Salpeter
Equation}\label{Sec:DB}For the spinless Salpeter equation, results
on the number of bound states are sparse \cite{Daubechies,Brau};
an upper limit on this quantity as easy to handle as the Bargmann
limit has been proved by I.~Daubechies \cite{Daubechies}: for
technical reasons, let the Hamiltonian operator $H=K+V$~acting on
$L^2({\mathbb R}^3)$ be composed of a kinetic term $K$ that is a
positive, strictly increasing, differentiable function of
$|\bm{p}|$ only that vanishes at $|\bm{p}|=0$ (which in the
relativistic case can be effected by subtracting appropriate
multiples of $m$) and rises beyond bounds with increasing
$|\bm{p}|$ and of some potential $V$ that is a negative smooth
function of compact support, $V\in C^\infty_0({\mathbb
R}^3),$~\emph{i.e.},$$H\equiv K(|\bm{p}|)+V(\bm{x})\
,\qquad\!K(|\bm{p}|)\ge0\ ,\qquad\!K(0)=0\
,\qquad\!K(|\bm{p}|)\xrightarrow[\bm{p}\to\infty]{}\infty\
,\qquad\!V(\bm{x})\le0\ .$$The total number $N$ of bound states of
such suitable operators is subject to the upper~limit
\begin{equation}N\le\frac{C}{6\,\pi^2}\int{\rm d}^3x
\left[K^{-1}(|V(\bm{x})|)\right]^3\ ,\label{Eq:DB}\end{equation}
which involves a constant $C$ depending on the kinetic energy $K$
but not on the potential~$V$:
$$C=\inf_{b>0}\!\left(\left\{e^b\int_0^\infty\frac{{\rm
d}y}{y^2}\,e^{-b\,y}\,[g(y)]^3\right\}
\left\{b\int_0^\infty\frac{{\rm
d}y\,y}{y+1}\,e^{-b\,y}\right\}^{\!-1}\right),\qquad
g(y)\equiv\sup_{x>0}\frac{K^{-1}(x\,y)}{K^{-1}(x)}\
.$$Specifically, subsuming one-body and equal-mass two-body
\emph{relativistic kinematics},~we~have\begin{equation}K(|\bm{p}|)
=\alpha\left(\sqrt{|\bm{p}|^2+m^2}-m\right),\qquad\alpha=1,2\
.\label{Eq:RKE}\end{equation}The constant $C$ is independent also
of $\alpha$: $C=6.074898$ for $m=0,$ $C=14.10759$~for $m>0.$ The
number of bound states of a spinless Salpeter equation with
potential $V(r)$~is,~at~most,$$N\le \frac{C}{6\,\pi^2}\int{\rm
d}^3x\left[\frac{|V(r)|}{\alpha}
\left(\frac{|V(r)|}{\alpha}+2\,m\right)\right]^{\!3/2}\ .$$In
particular, in the case of two bound particles of equal mass
($\alpha=2$), this inequality~reads$$N\le\frac{C}{48\,\pi^2}
\int{\rm d}^3x\left[|V(r)|\left(|V(r)|+4\,m\right)\right]^{3/2}
=\frac{C}{12\,\pi}\int_0^\infty{\rm d}r\,r^2
\left[|V(r)|\left(|V(r)|+4\,m\right)\right]^{3/2}\ .$$

\section{Schr\"odinger Upper Bounds on Energy Eigenvalues}
\label{Sec:SchrUB}Unspectacular upper bounds on the eigenvalues of
any semirelativistic Hamiltonian~(\ref{Eq:H})~are provided by the
eigenvalues of its nonrelativistic limit $H_{\rm NR}$
\cite{Lucha96:RCP,Lucha99:talksa,Lucha99:talksb,Lucha04:TWR},
since the positivity
$$\left(\sqrt{\bm{p}^2+m^2}-m\right)^{\!2}\ge0$$of the square of
the self-adjoint operator $\sqrt{\bm{p}^2+m^2}-m$ means that for
any spinless particle the relativistic kinetic energy is bounded
from above by the nonrelativistic kinetic
energy:\footnote{Geometrically, this is just a consequence of the
concavity of the square root. The nonrelativistic kinetic energy
is the tangent line to the square root of the relativistic kinetic
energy, at the point of~contact
$\bm{p}^2=0.$}$$\sqrt{\bm{p}^2+m^2}\le m+\frac{\bm{p}^2}{2\,m}\
.$$This operator inequality carries over to the full Hamiltonian
$H,$ involving a potential $V(\bm{x});$ the semirelativistic
operator is bounded from above by its nonrelativistic counterpart
$H_{\rm NR}$:$$H\equiv T(\bm{p})+V(\bm{x})\le H_{\rm
NR}\equiv2\,m+\frac{\bm{p}^2}{m}+V(\bm{x})\ .$$With this
inequality at one's disposal, it is easy to convince oneself
\cite{Lucha96:RCP,Lucha99:talksa,Lucha99:talksb,Lucha04:TWR}
that~any pair $(E_k,E_{k,{\rm NR}})$ of associated eigenvalues
necessarily satisfies $E_k\le E_{k,{\rm NR}}$ for all
$k=0,1,2,\dots.$

\section{Laguerre Variational Upper Bounds to Eigenvalues}
\label{Sec:LagVUP}The \emph{minimum--maximum theorem\/}
\cite{MMTa,MMTb,MMTc} offers easy-to-handle tools for the
localization of the \emph{discrete\/} spectrum of an operator $H$
in Hilbert space: Let $H$ be \emph{self-adjoint\/} and
\emph{bounded from below\/} and its \emph{eigenvalues\/} $E_k,$
defined by the eigenvalue equation
$H\,|\chi_k\rangle=E_k\,|\chi_k\rangle$ for the associated
eigenstates $|\chi_k\rangle,$ $k=0,1,\dots,$ be \emph{ordered\/}
according to $E_0\le E_1\le E_2\le\cdots.$ Then, restricting $H$
to a $d$-dimensional \emph{subspace\/} $D_d$ of its domain, the
eigenvalues $E_k$ \emph{below\/} the onset of the
\emph{essential\/} spectrum of $H$ if counting multiplicity of
degenerate states satisfy
$$E_k\le\displaystyle\sup_{|\psi\rangle\in D_{k+1}}
\frac{\langle\psi|\,H\,|\psi\rangle}{\langle\psi|\psi\rangle}
\qquad\mbox{for all}\ k=0,1,2,\dots\ .$$An immediate consequence
of this theorem is the Rayleigh--Ritz variational technique. The
$d$ eigenvalues $\widehat E_k,$ $k=0,1,\dots,d-1,$ of this
operator $H$ restricted to a subspace~$D_d,$~likewise ordered
according to $\widehat E_0\le\widehat E_1\le\dots\le\widehat
E_{d-1},$ are \emph{upper bounds\/} to the first $d$
eigenvalues~$E_k$:$$E_k\le\widehat E_k\ ,\qquad k=0,1,\dots,d-1\
.$$The quality or tightness \cite{Lucha:Q&Aa,Lucha:Q&Ab} of upper
bounds found in this way may be increased~by~just enlarging the
trial space $D_d$ by raising its dimension; at least, this won't
make things worse.

The Woods--Saxon potential (\ref{Eq:WSP}) possesses the merit of
being non-singular: the operator (\ref{Eq:H}) is bounded from
below. For a relativistic Coulomb problem, posed by assuming
$V(\bm{x})$ to be the Coulomb potential $V_{\rm
C}(\bm{x})\propto-1/|\bm{x}|,$ for instance, both essential
self-adjointness and boundedness from below of the operator
(\ref{Eq:H}) has been rigorously established by Herbst
\cite{Herbst}.

Hence, we feel entitled to exploit the minimum--maximum theorem
for the construction of rigorous upper bounds to the energy
eigenvalues $E_k$ of the semirelativistic Woods--Saxon
Hamiltonian. By spherical symmetry, every basis function of the
Hilbert space~$L^2({\mathbb R}^3)$ (\emph{i.e.}, the Hilbert space
of the square-integrable functions on the three-dimensional space
${\mathbb R}^3$) is a product of a radial function and a spherical
harmonic ${\cal Y}_{\ell m}(\Omega)$ for angular momentum $\ell$
and projection $m$ that depends on the solid angle $\Omega$ in the
underlying representation space. We span the trial spaces $D_d$ of
the minimum--maximum theorem by an orthonormal set of basis
vectors \cite{Jacobs86,Lucha97:LagBd,Lucha99:talksa,
Lucha99:talksb,Lucha04:TWR}, involving two variational parameters:
$\mu,$ with the dimension of mass, and $\beta,$ being
dimensionless; our basis states exhibit the compulsory
normalizability as long as the variational parameters assume
values in the intervals $0<\mu<\infty$
and~$-\frac{1}{2}<\beta<\infty.$\begin{itemize}\item The
\emph{configuration-space representation\/} $\psi_{k,\ell
m}(\bm{x})$ of our orthonormalized basis vectors utilizes the
\emph{generalized-Laguerre orthogonal polynomials\/}
$L_k^{(\gamma)}(x)$ for the parameter $\gamma,$ defined by a power
series \cite{Abramowitz,Bateman}, with orthogonality fixing
$\gamma$ uniquely to $\gamma=2\,\ell+2\,\beta$:$$\psi_{k,\ell
m}(\bm{x})=\sqrt{\frac{(2\,\mu)^{2\ell+2\beta+1}\,k!}
{\Gamma(2\,\ell+2\,\beta+k+1)}}\,|\bm{x}|^{\ell+\beta-1}
\exp(-\mu\,|\bm{x}|)\,L_k^{(2\ell+2\beta)}(2\,\mu\,|\bm{x}|)\,
{\cal Y}_{\ell m}(\Omega_{\bm{x}})\ ,$$ $$L_k^{(\gamma)}(x)\equiv
\sum_{t=0}^k\,(-1)^t\binom{k+\gamma}{k-t}\frac{x^t}{t!}\ ,\qquad
k=0,1,2,\dots\ ,$$ $$\int{\rm d}^3x\,\psi_{k,\ell
m}^\ast(\bm{x)}\,\psi_{k',\ell'm'}(\bm{x)}=
\delta_{kk'}\,\delta_{\ell\ell'}\,\delta_{mm'}\ .$$\item The
\emph{momentum-space representation\/} $\widetilde\psi_{k,\ell
m}(\bm{p})$ of the orthonormalized basis vectors is found to be a
series involving the hypergeometric function $F(u,v;w;z)$
\cite{Abramowitz}, which, in turn, may be defined, in terms of the
gamma function $\Gamma(z),$ by its Gauss power~series:
\begin{align*}\widetilde\psi_{k,\ell m}(\bm{p})&=
\sqrt{\frac{(2\,\mu)^{2\ell+2\beta+1}\,k!}
{\Gamma(2\,\ell+2\,\beta+k+1)}}\,\frac{(-{\rm
i})^\ell\,|\bm{p}|^\ell}{2^{\ell+1/2}\,
\Gamma\!\left(\ell+\frac{3}{2}\right)}\\[1ex]&\times
\sum_{t=0}^k\,\frac{(-1)^t}{t!}\binom{k+2\,\ell+2\,\beta}{k-t}
\frac{\Gamma(2\,\ell+\beta+t+2)\,(2\,\mu)^t}
{(\bm{p}^2+\mu^2)^{(2\ell+\beta+t+2)/2}}\\[1ex]&\times
F\!\left(\frac{2\,\ell+\beta+t+2}{2},-\frac{\beta+t}{2};
\ell+\frac{3}{2};\frac{\bm{p}^2}{\bm{p}^2+\mu^2}\right){\cal
Y}_{\ell m}(\Omega_{\bm{p}})\ ,\end{align*}$$F(u,v;w;z)\equiv
\frac{\Gamma(w)}{\Gamma(u)\,\Gamma(v)}\,\sum_{n=0}^\infty\,
\frac{\Gamma(u+n)\,\Gamma(v+n)}{\Gamma(w+n)}\,\frac{z^n}{n!}\ ,$$
$$\int{\rm d}^3p\,\widetilde\psi_{k,\ell
m}^\ast(\bm{p)}\,\widetilde\psi_{k',\ell'm'}(\bm{p)}=
\delta_{kk'}\,\delta_{\ell\ell'}\,\delta_{mm'}\ .$$\end{itemize}
Of course, for any basis state its configuration-space and
momentum-space representations $\psi_{k,\ell m}(\bm{x})$ and
$\widetilde\psi_{k,\ell m}(\bm{p})$ are related by Fourier
transformations acting on $L^2({\mathbb R}^3),$ their radial
functions consequently by Fourier--Bessel transformations. The
advantage of such~choice of basis is the availability of both
representations in \emph{analytic\/} form. This enables us to
evaluate the matrix elements of $T(\bm{p})$ in momentum space, and
those of $V(\bm{x})$ in configuration space.

\section{Approximations Enabling an Analytical Treatment}
\label{Sec:pSSE}More or less recently, a hardly manageable amount
of investigations aiming at approximate but to the largest
reasonable extent analytical approaches to the spinless Salpeter
equation have been published. Driven by the maybe easily
comprehensible desire to achieve this goal by all available means,
some of such analyses feel forced to refer to disputable
assumptions.

The first of these approximations changes the nature of this
problem entirely. Assuming sufficiently heavy bound-state
constituents, the free energy is expanded nonrelativistically:
$$\sqrt{\bm{p}^2+m^2}\approx m+\frac{\bm{p}^2}{2\,m}
-\frac{\bm{p}^4}{8\,m^3}+\cdots.$$ Disregarding in this expansion
terms of higher than second order in $\bm{p}^2/m^2,$ \emph{i.e.},
truncating this power series at the order $\bm{p}^4/m^4,$
generates the ``pseudo spinless-Salpeter Hamiltonian''
\begin{equation}H_{\rm p}\equiv2\,m+\frac{\bm{p}^2}{m}
-\frac{\bm{p}^4}{4\,m^3}+V(\bm{x})\ .\label{Eq:Hp}\end{equation}
However, leaving aside the extremely unlikely possibility of a
kind of mysterious conspiracy of the involved potential
$V(\bm{x}),$ this Hamiltonian $H_{\rm p},$ as an operator on
$L^2({\mathbb R}^3),$ is definitely unbounded from below; as a
consequence, a ground state of $H_{\rm p}$ does not exist. This is
easily seen by considering the expectation value of $H_{\rm p}$
with respect to suitably chosen trial states; for instance, for
the Laguerre basis state of Sec.~\ref{Sec:LagVUP} for the quantum
numbers~$n=\ell=m=0,$$$\psi_{0,00}(\bm{x})=\sqrt{\frac{\mu^3}{\pi}}
\exp(-\mu\,|\bm{x}|)\ ,\qquad\widetilde\psi_{0,00}(\bm{p})=
\frac{\sqrt{8\,\mu^5}}{\pi}\,\frac{1}{(\bm{p}^2+\mu^2)^2}\
,\qquad\beta=1 \ ,$$the resulting expectation value $\langle
H_{\rm p}\rangle$ behaves as function of the variational
parameter~$\mu$~like$$\langle H_{\rm
p}\rangle=2\,m+\frac{\mu^2}{m} -\frac{5\,\mu^4}{4\,m^3}+\langle
V(\bm{x})\rangle\qquad\Longrightarrow\qquad\lim_{\mu\to\infty}\langle
H_{\rm p}\rangle=-\infty\qquad\Longrightarrow\qquad E_0\le-\infty\
.$$Clearly, such annoying negligibility does not show up if, for
some reasons, the term of~order $\bm{p}^4/m^4$ enters into one's
bound-state equation with the wrong sign \cite[Eq.~(2)]{IS_ZPC92}
\cite[Eq.~(2)]{IS_ZPC93}. If, however, the $\bm{p}^4/m^4$ term
retains its correct sign, the eigenvalue equation of the operator
$H_{\rm p}$ \cite[Eq.~(3)]{IS_IJMPA04} \cite[Eq.~(3)]{IS_IJMPA05}
\cite[Eq.~(13)]{IS_IJMPE08-2} \cite[Eq.~(4)]{Feizi_IJMPE} is to be
taken with a big grain~of~salt. A promising remedy is the
perturbative treatment of such bothersome contributions~to~$H_{\rm
p}.$

In order to eventually enforce \emph{analytical solvability\/} of
one's \emph{radial Schr\"odinger equation}, its centrifugal term
with orbital quantum number $\ell,$ $\ell\,(\ell+1)/r^2,$ arising
from the Laplacian, is approximated \cite{Pekeris} by an expansion
in terms of a function $y(r)$ resembling one's
potential:$$\frac{1}{r^2}\stackrel{!}{\approx}f(r)\equiv
\sum_{j=0}^{P<\infty}c_j\,[y(r)]^j\ ,\qquad P\ge2\ .$$Thus, in our
case the function of choice for the new variable $y(r)$ is the
Woods--Saxon~shape
$$y(r)\equiv\frac{1}{1+\exp\!\left(\frac{r-R}{a}\right)}\ .$$
However, the behaviour of $f(r)$ differs drastically from that of
the centrifugal term; besides regularising the $1/r^2$ singularity
at $r=0,$ it will \emph{not\/} unavoidably approach zero for
$r\to\infty$:$$\lim_{r\to\infty}y(r)=0\qquad\Longrightarrow\qquad
\lim_{r\to\infty}f(r)=c_0<\infty\ .$$In contrast to any
\emph{actual\/} Woods--Saxon problem, for $\ell\ne0$ this allows
also for bound states with (fake) \emph{positive\/} energy
eigenvalues $E^{\rm f}_k$ that are bounded from above by $E^{\rm
f}_k\le\ell\,(\ell+1)\,c_0.$

\section{Properties of Solutions to Woods--Saxon Problems}Hence,
the way is paved for sketching our picture of the relativistic
Woods--Saxon problem.

One of the prerequisites for the upper bound (\ref{Eq:DB}) to hold
for generic kinetic terms $K(|\bm{p}|)$ (provided the latter
comply with all assumptions necessary for proving this relation)
is that the potential $V(\bm{x})$ is a smooth function with
compact support, $V\in C^\infty_0({\mathbb R}^3).$ However, the
Woods--Saxon potential (\ref{Eq:WSP}) does not have compact
support and thus does~not belong to the function space
$C^\infty_0({\mathbb R}^3).$ Fortunately, it can be shown
\cite{Daubechies} that, for relativistic~kinematics as represented
by the kinetic-energy operator (\ref{Eq:RKE}), the range of
validity of the upper bound (\ref{Eq:DB}) extends to all
potentials $V(\bm{x})$ in $L^{3/2}({\mathbb R}^3)\cap L^3({\mathbb
R}^3).$ Since the Woods--Saxon potential~(\ref{Eq:WSP}) is --- for
finite values of all three parameters (\emph{i.e.}, $V_0,$ $R$ and
$a$) characterizing this potential~--- in this class, we expect to
be on the safe side when applying Eq.~(\ref{Eq:DB}) to our
Hamiltonian~(\ref{Eq:H}).

A trivial lower bound to the spectrum $\sigma(H)$ of any
Hamiltonian $H$ of the shape~(\ref{Eq:H0}) with positive
kinetic-energy operator $T(\bm{p})\ge0$ is provided by the infimum
of its potential~$V(\bm{x})$:$$\sigma(H)\ge\inf_{\bm{x}}V(\bm{x})\
.$$For the Woods--Saxon potential (\ref{Eq:WSP}), this lower
energy bound is, clearly, realized in the~form
\begin{equation}\sigma(H)\ge\inf_rV(r)=\min_rV(r)=V(0)
=-\frac{V_0}{1+\exp\!\left(-\frac{R}{a}\right)}\ge-V_0\
.\label{Eq:LSB}\end{equation}

We are aware of three studies \cite{Hamzavi_CPC,Hassanabadi_CPC,
Feizi_IJMPE} attempting to construct approximate solutions, by the
analytic path recalled in Sec.~\ref{Sec:pSSE}, to the spinless
Salpeter equation with Woods--Saxon potential. Out of these,
Ref.~\cite{Hamzavi_CPC} provides its results in a most detailed
and explicit manner. Consequently, let us present the outcome of
our considerations too for the numerical values adopted in the
investigation of Ref.~\cite{Hamzavi_CPC} for the mass $m$ of both
of the involved bound-state constituents and the parameters $V_0,$
$R,$ and $a$ determining the real Woods--Saxon~potential:
\begin{align*}m&=4.76504\;\mbox{fm}^{-1}=0.940271\;\mbox{GeV}\ ,\\
V_0&=0.3431032\;\mbox{fm}^{-1}=0.06770352\;\mbox{GeV}\ ,\\
R&=7.6136\;\mbox{fm}=38.584\;\mbox{GeV}^{-1}\ ,\\
a&=0.65\;\mbox{fm}=3.3\;\mbox{GeV}^{-1}\ .\end{align*}For this set
of parameters, we compute the \emph{upper limits\/} on the number
of bound states that can be accommodated by the particular
potential specified by this choice according to both
nonrelativistic and relativistic kinematics discussed in
Sec.~\ref{Sec:NBS}, the variational upper bounds on the energy
eigenvalues of the \emph{spinless Salpeter equation\/}, generated
by the Rayleigh--Ritz technique summarized in
Sec.~\ref{Sec:LagVUP}, and (with the help of an easily applicable
standard routine for a straightforward numerical solution of
Schr\"odinger equations \cite{Lucha98}) their nonrelativistic
counterparts reproduced in Sec.~\ref{Sec:SchrUB}. The lower energy
bound (\ref{Eq:LSB}) applying likewise to spinless Salpeter and
Schr\"odinger cases is found instantly: $E_0\ge
V(0)=-0.06770296\;\mbox{GeV}\ge-V_0.$

For both nonrelativistic and relativistic spinless Woods--Saxon
problems, the maximum number $N$ of bound states we may encounter
is, trivially, \emph{finite\/}: $0\le N<\infty.$ For the choice
used in Ref.~\cite{Hamzavi_CPC} for the mass and potential
parameters, we obtain for the number~$N,$ by the Daubechies bound
in Eq.~(\ref{Eq:DB}), $N\le850$ and, by the Bargmann bound in
Eq.~(\ref{Eq:BB}), $N\le1201.$ So, for the present parameters, one
has enough Woods--Saxon bound states~to~play around.

Table \ref{Tab:RWSP} presents our variational and Schr\"odinger
upper limits to the \emph{binding energies\/} of the lowest
Woods--Saxon bound states with orbital angular momentum quantum
number $\ell$ and radial quantum number $n_r,$ defined as the
number of zeros of the radial wave~functions.

\newpage

\begin{table}[ht]\caption{Upper limits to the binding energy for
the lowest-lying bound states of the spinless Salpeter equation
with real Woods--Saxon potential, for the set of parameter values
used in Ref.~\cite{Hamzavi_CPC}: the trivial Schr\"odinger bounds
$\overline{E}_{\rm NR}$ of Sec.~\ref{Sec:SchrUB} and the Laguerre
bounds $\overline{E}$~of~Sec.~\ref{Sec:LagVUP}. Any bound state is
identified by its radial ($n_r$) and orbital angular momentum
($\ell$) quantum numbers. For the bound-state problem investigated
here, optimization of the orthogonality of the resulting
eigenstates fixes the dimension $d$ of the variational trial space
$D_d$ to $d=25$; as a mere illustration, the variational
parameters $\mu$ and $\beta$ are kept fixed, $\mu=1\;\mbox{GeV},$
$\beta=1.$}\label{Tab:RWSP}
\begin{center}\begin{tabular}{cccc}\hline\hline\\[-1.5ex]
\multicolumn{2}{c}{Bound state}&\multicolumn{1}{c}{Spinless
Salpeter equation}&\multicolumn{1}{c}{Schr\"odinger
equation}\\[1.5ex]\cline{1-2}\\[-1.5ex]$n_r$&$\ell$&
\multicolumn{1}{c}{$\overline{E}(n_r,\ell)\;[\mbox{GeV}]$}&
\multicolumn{1}{c}{$\overline{E}_{\rm
NR}(n_r,\ell)\;[\mbox{GeV}]$}\\[1.5ex]\hline\\[-1.5ex]
0&0&$-0.06032$&$-0.06030$\\&1&$-0.05309$&$-0.05305$\\[1ex]
1&0&$-0.04119$&$-0.04108$\\&1&$-0.02967$&$-0.02946$\\[1ex]
2&0&$-0.01527$&$-0.01545$\\&1&$-0.00233$&$-0.00362$
\\[1.5ex]\hline\hline\end{tabular}\end{center}\end{table}

The main assumption underlying every search for approximate
solutions to the spinless Salpeter equation based on a
\emph{nonrelativistic\/} expansion of the relativistic free energy
[which yields, as the eigenvalue equation of the Hamiltonian
(\ref{Eq:Hp}), a kind of pseudo spinless Salpeter equation] is
that, for the given potential, all bound-state constituents are
sufficiently heavy for this drastic modification to make sense.
Before embarking on the detailed comparison~of our insights with
the findings of Ref.~\cite{Hamzavi_CPC}, let us estimate the
extent to which this expansion is justified. Under the reasonable
assumption that the lowest-lying bound state deduced~by
application of the variational technique sketched in
Sec.~\ref{Sec:LagVUP}, with radial and orbital quantum numbers
$n_r=\ell=0,$ represents a passable description of the ground
state of the relativistic Woods--Saxon problem, the expectation
value of the first non-trivial term in this expansion of the
operator $\sqrt{\bm{p}^2+m^2}$ divided by the mass $m,$ taken with
respect to this bound state,~is
$$\left\langle\frac{\bm{p}^2}{m^2}\right\rangle=6.2\times10^{-3}\
.$$Thus, the assumption is indeed justified: for the parameter
values of Ref.~\cite{Hamzavi_CPC}, the system is highly
nonrelativistic. As a matter of fact, a question that inevitably
comes to one's mind is whether, \emph{for such systems}, it is
worthwhile to invest all the efforts to overcome all obstacles of
some pseudo spinless Salpeter treatment instead of sticking to the
Schr\"odinger~equation.

We take the liberty of regarding that coarse characterization of
the eigenvalue spectrum of the relativistic Woods--Saxon problem
given above as a collection of rigorous \emph{constraints\/} that
any solution of the spinless Salpeter equation with Woods--Saxon
potential must fulfil. The latter request should constitute for,
of course, all proposed solutions, but especially for approximate
solutions of unclear quality or accuracy, a crucial criterion for
their reliability. Following this spirit, let us look how well
this overall pattern is reflected in
Refs.~\cite{Hamzavi_CPC,Hassanabadi_CPC,Feizi_IJMPE}.

However, even a merely cursory inspection of the claims of
Ref.~\cite{Hamzavi_CPC} leads to a number~of observations which
cast serious doubts on the reliability of the results reported by
Ref.~\cite{Hamzavi_CPC}:\begin{itemize}\item To begin with, one's
confidence in all the assertions of Ref.~\cite{Hamzavi_CPC} would
be significantly strengthened if the binding-energy eigenvalues of
the (only perturbatively accessible) pseudo spinless-Salpeter
Hamiltonian defined in Eq.~(\ref{Eq:Hp}) \cite[Table
1]{Hamzavi_CPC} and those~of the corresponding Schr\"odinger
Woods--Saxon problem \cite[Table 2]{Hamzavi_CPC} would be
compatible with our knowledge about crucial overall features of
the spectrum of the Hamiltonian (\ref{Eq:H}) with Woods--Saxon
potential and its nonrelativistic counterpart. Regrettably, the
authors of Ref.~\cite{Hamzavi_CPC} preferred not to reveal the
units of the binding-energy eigenvalues collected in their Tables
1 and 2. However, they present the adopted numerical~values of
their parameters in units of appropriate powers of fermi. So, in
order to proceed we assume that their binding-energy eigenvalues
too are quoted in units of inverse fermi. To our great amazement,
we then realize that \emph{all\/} binding-energy eigenvalues
listed in Table 1 of Ref.~\cite{Hamzavi_CPC} and a considerable
fraction of binding-energy eigenvalues~listed in Table 2 of
Ref.~\cite{Hamzavi_CPC} lie \emph{below\/} our trivial lower
spectral bound (\ref{Eq:LSB}), and hence~cannot~be part of the
spectra of the corresponding Hamiltonians. We regard this as a
convincing indication that the entire content of the tables in
Ref.~\cite{Hamzavi_CPC} is rather far from the~truth.\item
Explicit analytic expressions for the binding-energy eigenvalues
of the nonrelativistic Schr\"odinger equation with Woods--Saxon
potential (\ref{Eq:WSP}) can be found twice in
Ref.~\cite{Hamzavi_CPC} --- relations (23) and (26) in
Ref.~\cite{Hamzavi_CPC} are, in fact, \emph{identical\/}. However,
this duplication \cite[Eq.~(23)]{Hamzavi_CPC},
\cite[Eq.~(26)]{Hamzavi_CPC} cannot hide the fact that already on
dimensional grounds this expression can by no means be correct: a
quantity with non-zero mass dimension is obviously missing in the
term which involves the quantum-number difference $n-L.$
Nevertheless, the incorrect expression has apparently been adopted
by the authors of Ref.~\cite{Hamzavi_CPC} for computing the
entries of their Table 2, rendering the table rather useless. The
--- within the plethora of approximations made in
Ref.~\cite{Hamzavi_CPC} --- true expression for the
nonrelativistic binding energies reads, for vanishing
orbital~angular~momentum~$\ell,$\begin{equation}E_{n,0}=-\frac{1}{m}
\left(\frac{m\,V_0\,a^2+n^2}{2\,n\,a}\right)^{\!2}\qquad\mbox{for}
\quad m\equiv m_1=m_2\ .\label{Eq:En0}\end{equation}\item In
quantum theory, a preeminent criterion for the reliability of
solutions to eigenvalue equations of self-adjoint operators is
established by a generalization \cite{Lucha:RVTb}, to arbitrary
kinematics, of the virial theorem well known from nonrelativistic
quantum mechanics \cite{VTQMa,VTQMb,VTQMc}. For an operator $H$ of
the form (\ref{Eq:H0}), this master virial theorem
\cite{Lucha:RVTb} claims that the expectation values, taken with
respect to a given eigenstate $|\chi\rangle$ of the operator $H,$
of the (momentum-space) \emph{radial\/} derivative of its
kinetic-energy operator $T(\bm{p})$ and of the
(configuration-space) \emph{radial\/} derivative of its
interaction potential~$V(\bm{x})$~are~equal:$$\left\langle\chi
\left|\,\bm{p}\cdot\frac{\partial\,T}{\partial\bm{p}}(\bm{p})
\,\right|\chi\right\rangle=\left\langle\chi\left|\,\bm{x}\cdot
\frac{\partial\,V}{\partial\bm{x}}(\bm{x})\,\right|\chi
\right\rangle.$$This (universal) relation can be straightforwardly
employed for checking, one by one, presumed eigenstates, or
approximations to the latter, for their validity. For~instance,
the bound state associated to the lowest of all binding-energy
upper limits in Table~\ref{Tab:RWSP}, consequently approximating
the ground state, \emph{i.e.}, $n_r=\ell=0,$ of the
semirelativistic spinless Woods--Saxon problem investigated here,
passes this test with flying colours:$$\left\langle
\frac{2\,\bm{p}^2}{\sqrt{\bm{p}^2+m^2}}\right\rangle
=0.0116\;\mbox{GeV}=\left\langle r\,\frac{\partial\,V}{\partial
r}(r)\right\rangle.$$Sadly, even the approximate Schr\"odinger
state for the quantum numbers $n=1,\ell=0$ determined by
Eqs.~(24), (25) of Ref.~\cite{Hamzavi_CPC} for the binding energy
$E_{1,0}=-0.0700\;\mbox{GeV}$ computed from the corrected
expression (\ref{Eq:En0}) misses this not too ambitious goal
badly:$$\left\langle\frac{2\,\bm{p}^2}{m}\right\rangle
=0.0379\;\mbox{GeV}\ne0.0415\;\mbox{GeV}=\left\langle
r\,\frac{\partial\,V}{\partial r}(r)\right\rangle.$$\item Section
3 of Ref.~\cite{Hamzavi_CPC} offers, at the price of a certain
lack of rigour, an analytical albeit approximate solution
\cite[Eqs.~(22)--(25)]{Hamzavi_CPC} to the Schr\"odinger
Woods--Saxon problem, deduced as the nonrelativistic limit of
corresponding results \cite[Eqs.~(15), (16), (20)]{Hamzavi_CPC}
emerging, in Sec.~2 of Ref.~\cite{Hamzavi_CPC}, from an
unavoidably merely perturbative approach to any bound-state
problem posed by the pseudo spinless-Salpeter Hamiltonian
(\ref{Eq:Hp}). To seek an analytical discussion of a given physics
problem is, of course, a legitimate goal and the successful
completion of such task even a highly desirable feature of
proposed solutions but has to be backed by numerical verification.
Particularly for Schr\"odinger operators a huge variety of results
on their spectral decomposition has been compiled and has found
its way into the textbooks. Such findings provide straightforward
tools for checking the self-consistency of alleged analytical
solutions; numerical procedures for solving the Schr\"odinger
equation are abundant in the literature, cf., \emph{e.g.},
Ref.~\cite{Lucha98}. A lot of considerations can be performed
without entering or even touching the realm of mathematical
physics. Let us illustrate this for the ($n=1,\ell=0$) state
represented by Eqs.~(27), (28) of Ref.~\cite{Hamzavi_CPC}, with
binding energy $E_{1,0}=-0.0700\;\mbox{GeV}$~from
Eq.~(\ref{Eq:En0}):\begin{itemize}\item The \emph{expectation
value\/} of a given operator $H$ in Hilbert space with respect to
any of its eigenstates $|\chi_k\rangle$ is necessarily identical
to its corresponding \emph{eigenvalue\/}~$E_k$:
$$H\,|\chi_k\rangle=E_k\,|\chi_k\rangle\qquad\Longrightarrow\qquad
\frac{\langle\chi_k|\,H\,|\chi_k\rangle}{\langle\chi_k|\chi_k\rangle}
=E_k\ ,\qquad k=0,1,2,\dots\ .$$In the meantime without surprise,
such triviality is not satisfied by our test~case:
$$\left\langle\frac{\bm{p}^2}{m}+V(r)\right\rangle
=-0.0377\;\mbox{GeV}\ne-0.0700\;\mbox{GeV}=E_{1,0}\
.$$Irrespective of an interpretation of Eq.~(21) of
Ref.~\cite{Hamzavi_CPC} from the physics point of view as a
quantum-mechanical Schr\"odinger equation, Eqs.~(23)--(25) of
Ref.~\cite{Hamzavi_CPC} provide a rather poor approximation to the
solution of this differential equation.\item Even if leaving aside
the restrictions imposed by requiring self-adjointness of the
Schr\"odinger operator $H$ on $L^2({\mathbb R}^3)$ whose
eigenvalue equation yields the \emph{reduced\/} radial
Schr\"odinger equation recalled in Ref.~\cite{Hamzavi_CPC} as
Eq.~(21), simple inspection of the behaviour of the position-space
\emph{reduced\/} radial wave functions $\psi_{n,\ell}(r)$ solving
the bound-state equation forces all its solutions to vanish
at~spatial~origin $r=0$:$$\lim_{r\to0}\psi_{n,\ell}(r)\stackrel{!}
{=}0\ .$$Easy to guess, this property is not shared by all
solutions in Eq.~(24) of Ref.~\cite{Hamzavi_CPC}:$$|\psi_{1,0}(0)|
=0.06879\;\mbox{GeV}^{1/2}\ne0\ .$$\end{itemize}As the
Schr\"odinger case forms the nonrelativistic endpoint of the
analysis in Ref.~\cite{Hamzavi_CPC}, all this does not lend too
much trust in the results of the procedures used by
Ref.~\cite{Hamzavi_CPC}.\end{itemize}

In Ref.~\cite{Hassanabadi_CPC}, the emerging energy eigenvalues
are left encoded in an implicit relation and still await their
numerical disclosure \cite[Eq.~(19)]{Hassanabadi_CPC}; hence,
there is nothing to compare~with.

\newpage

In Ref.~\cite{Feizi_IJMPE}, numerical values of the binding
energies of a few bound states are explicitly given for three
different sets of mass and potential parameters \cite[Table
1]{Feizi_IJMPE}. For all of~these sets, the expectation values of
$\bm{p}^2/m^2$ over our variational ground state fall into the
interval$$6.9\times10^{-3}\le
\left\langle\frac{\bm{p}^2}{m^2}\right\rangle\le9.3\times10^{-2}\
;$$so, all three systems are also sufficiently nonrelativistic to
provide some justification for the expansion of that square root
of the relativistic kinetic-energy operator in powers of
$\bm{p}^2/m^2.$ All the binding-energy values provided in
Ref.~\cite{Feizi_IJMPE} are larger than the lower~spectral bound
(\ref{Eq:LSB}), $V(0)\gtrapprox-V_0.$ The question whether further
states satisfy this bound is left~unanswered.

\section{Summary and Conclusions}As a kind of comment on the
current quest for an analytic approach to the spinless Salpeter
equation, we compiled in this brief note some general statements
of quantum theory related to any such enterprise and applied these
insights to the case of the Woods--Saxon potential. Although such
techniques do not immediately lead to the exact solutions, they
nevertheless constitute invaluable tools to scrutinize approximate
solutions for their significance. In this way, these (not
excessively complicated) considerations may be exploited to
systematically separate wheat from chaff, as has been demonstrated
for the findings presented in Ref.~\cite{Hamzavi_CPC}.

\end{document}